\documentclass[11pt,onecolumn]{article}
\usepackage{amsmath}

\usepackage{stmaryrd}
\usepackage {amsfonts}
\usepackage {amsmath,amssymb,amsthm}
\usepackage {txfonts}
\usepackage{times}
\usepackage[top=2.5cm, bottom=1cm, left=2.1cm, right=2.6cm]{geometry}
\usepackage{enumerate}

\newtheorem{definition}{Definition}

\begin{document}

\title{\Large{\textbf{A Calculus of Consistent Component-based Software Updates}}}
\author{Xiaohui Xu\quad Linpeng Huang\quad Dejun Wang\quad Junqing Chen \\ \\ Department of
Computer Science and Engineering,\\Shanghai Jiao Tong University,
Shanghai, 200240, China\\\emph{\{xuxiaohui, lphuang, wangdejun,
junqing2007\}@sjtu.edu.cn}}

\date{}
\pagestyle{empty}

\maketitle

\thispagestyle{empty}

\begin{abstract}

It is important to enable reasoning about the meaning and possible
effects of updates to ensure that the updated system operates
correctly. A formal, mathematical model of dynamic update should be
developed, in order to understand by both users and implementors of
update technology what design choices can be considered. In this
paper, we define a formal calculus $update\pi$, a variant extension
of higher-order $\pi$ calculus, to model dynamic updates of
component-based software, which is language and technology
independent. The calculus focuses on following main concepts: proper
granularity of update, timing of dynamic update, state
transformation between versions, update failure check and recovery.
We describe a series of rule on safe component updates to model some
general processes of dynamic update and discuss its reduction
semantics coincides with a labelled transition system semantics that
illustrate the expressive power of these calculi.

\end{abstract}


\section{Introduction}

Dynamic update\cite{Hicks05, Neamtiu09} is a general, software-based
technique: there is no need for redundant hardware or
special-purpose software architecture, and application state can be
naturally preserved between updated versions, so that current
processing is not compromised or interrupted. Most current works
about software update(e.g. \cite{Ajmani04, Hicks05, Neamtiu08,
Neamtiu09, Subramanian09}) concentrate on techniques and
implementations of bug-fixes or performance enhancements. They
implement some well-established update methods for global
synchronization of local updates or distributed updates. But they
lack both simplicity and generality and it is not clear what
properties are actually guaranteed. Therefore, we believe a formal,
mathematical model of dynamic update should be developed, in order
to understand what design choices can be considered, and what impact
they have on update complexity and scalability. There a few
formalization works have been carried out to support the correctness
and consistency derivations of dynamic update (e.g. \cite{Bierman03,
Sewell08, Stoyle07}. Nevertheless, we are not aware of past efforts
that have well formalized the models for timing of update, state
transformation, and update failure recovery.

To be able to reason and ensure properties about system
specifications supporting dynamic component updates, we need
mathematical tools. Process calculi are one suitable tool, providing
not only a description language, but a rigorous semantics as well,
allowing the proof of relevant properties. In this paper, we define
a formal calculus $update\pi$ to model dynamic updates of
component-based software, which is language and technology
independent. This calculus is an extension of the asynchronous,
polyadic and higher-order $\pi$-calculus based on the fact that
dynamic updates can be seen as interactive behaviors,
message-passing asynchronously, within a component-based system.
Some important aspects on dynamic component update, which ranging
from proper granularity of update, timing of dynamic update, state
preservation and transformation, and update failure check and
recovery, will be modeled and reduced in our calculus.

The rest of this paper is organized as follows. In Section 2, we
review the main considerations behind different constructs of the
$update\pi$ calculus, and then present the syntactical and
structural descriptions of $update\pi$ calculus. We describe in
Section 3 a series of rule on safe component updates to model some
general processes of dynamic update and some well-defined update
mechanism. The reduction semantics of the $update\pi$ calculus are
defined in Section 4, and show that the reduction semantics
coincides with a labelled transition system semantics, under
sufficient conditions on the interaction of processes with the
environment, that illustrate the expressive power of this calculi.
In section 5, we discuss related works on mechanisms and formalism
of dynamic updates. Section 6 concludes the paper with a discussion
of further work.

\section{The $update\pi$ Calculus}

In this section, we firstly motivate some design choices by
introducing informally the main elements about a dynamic component
update and the main constructs of the $update\pi$ calculus. Then we
present the syntactical and structural descriptions of $update\pi$
calculus, as a calculus for modeling dynamic updates of
component-based software, with a location concept to model the
hierarchical composition of components, a process sequence to model
the timing of updates, an accompanied process to model executed log
of updates, and a state increment or decrement mechanism to model
state preservation and transformation.

\subsection{Design Choices}

The $update\pi$ calculus inherits ideas from numerous previous
studies, included safe and timely dynamic updates\cite{Neamtiu09,
Stoyle07}, and higher-order calculus(specially, Kell calculus
\cite{Schmitt05}). It is built as an extension of the higher-order
$\pi$-calculus, and observes several main design principles which we
consider important for a foundational model of dynamic component
update: trusted update sources, reasonable timing of update, and
consistent state transformation.

In $update\pi$ calculus, exactly as in Kell calculus, we use the
hierarchical and programmable locality concept as a primitive form
of component that can be used simultaneously as a unit of
modularity, of isolation, and of passivation(the ability to freeze
and marshal a component during its execution). The $update\pi$
calculus is in fact a family of higher-order process calculi with a
special update locality, which shares the same basic operational
semantics rules with the other calculus, but differ in the language
used to define specified update mechanisms in input and output
constructs, and affiliate a state description for each process.
Furthermore, contrarily to some existing proposals, we interpret the
names declared inside a locality as private resources, that should
remain local to that locality. Neither do we include an axiom of
form $l[(\nu n) P] \equiv (\nu n) l[P]$ in structural congruence,
nor do we implement name extrusion across locality boundaries along
reduction steps that would require it(similar to some existing
implementations of $\pi$-calculus-related process
algebras\cite{Fournet02}). This design choices avoid ambiguous
diversity of reduction pathes in located process $l[P]$, and enable
safety of programmable locality where the resources are not
extravasate outside their owner.

The timing of an update, as many past researchers have observed, is
critical to ensuring the validity\cite{Gupta96, Hicks05}. This
synchronous update primitive dictates when an update can occur,and
makes it easier to understand the states of program which an update
is applied to than the alternative asynchronous approach, in which
an update could occur at any time. Some researches\cite{Hicks05,
Neamtiu08} have proved that synchronous updating makes it easier to
write correct updates. So some safe update points specified by a
programmer or implementor of updated software can express explicitly
the update requirements and favor to derive safe and timely software
updates. In our calculus, the process sequence $P\ ;Q$ forces a
temporal order between the two operands: process $Q$ will be
activated only after successful completion of $P$. Through the
process sequence, safe update points can be set in feasible
locations of whole programs, thus the timing of update is enforced
by some temporal operands of the sequential composition operator.

Without care, after several updates the state of an updated system
can become confusing, particularly when updates are in terms of
binary patches. Replacements and transformations must not interfere
with application access to the objects, and must be performed
efficiently in both space and time. However, the state to be
transformed might be corrupted (e.g., update for bug fixes) and
detecting and transforming a corrupted state to a non-corrupted
state is the quintessential state transformation problem. For
simplicity and without loss of generality, in this paper, we assume
that the state is not corrupted and the transformation can be done
safely so that important consistency is not broken. We propose an
approach to associate a state information to each process, where all
instant output actions of a process are recorded. In our calculus, a
related state $\delta$ is defined for each process. The process
state is composed of some output actions generated after execution
of the process, which denote the results of execution and are
expressed through the names as channels or locations.

\subsection{The Syntax}

The syntax of $update\pi$ calculus is given in Figure \ref{syntax}.
Similar to standard $\pi$-calculus processes, we use $\mathsf{0}$ to
denote a null process that does not perform any action, $X$ to
denote a unknown process which is waiting for an assignment, $P\ |\
Q$ to denote the parallel composition of two processes to allow
processes to interact, and $(\nu n)P$ to denote the name
restriction, in which the creation of a fresh name $n$ whose initial
scope is the process $P$. The process sequence $P\ ;Q$ forces a
temporal order between the two operands: process $Q$ will be active
only after successful completion of $P$. In this process sequence, a
sequence operator $;$ split two processes, where process $P$ is
assigned as a predecessor of the operator $;$ and then process $Q$
is a successor of this operator.

The interfaces of component are modeled by channels, thus the output
and input behaviors can be expressed through some actions on
channels. The channels can carry extensible records, which model
message exchanges between components through input and output
interfaces. An output on channel $a$ is noted $\bar{a}\langle
\omega\rangle$, where $\omega$ is a constant argument that can ba a
name $n$ or a process $P$. Specially, no continuation is to
afflicted each output action because we consider the output process
is parallel to other processes. Similar to Kell calculus, we use the
located process concepts, by the term $l[P]$, to model software
components for hierarchical composition. In the term $l[P]$, $l$ is
the name of the locality, that is, a identifier of location in which
the components are stored before the composition, and $P$ is the
process identifier of a single process, a process composition or a
process sequence executing at location $l$.

It is worth of notice that in an input trigger $\xi \ast P$, the
operator between input pattern $\xi$ and continuation process $P$ is
$\ast$ rather than $\cdot$, used in standard $\pi$-calculus. There
exist two cases for this operator: $\rhd$ and $\diamond$, where the
former is a disposable trigger operated only once and however the
latter can be preserved during a reduction. Our calculus implement
the primitive for recursion or replication through the $\diamond$
operator, which can model asynchronous message passing between
concurrent components. For example, the process $\bar{a}\langle
Q\rangle\ |\ a(X) \diamond P$ will be induced to $a(X) \diamond P\
|\ P\{Q/X\}$, where $\{Q/X\}$ denotes a capture avoiding
substitution of process variable $X$ with process $Q$. $a(x) * P$,
$a(X) * P$ respectively stands for a process willing to acquire a
resource: this can mean either receiving a first-order or
higher-order message. The input prefix $\xi$ and restriction
operator $\nu$ act respectively as a binder in the calculus. We
write $\mathrm{fn}(P)$ and $\mathrm{fv}(P)$ respectively for the
free names and free variables of process $P$. Furthermore, we use
the standard notions of free names of processes and of
$\alpha$-equivalence. We note $P=_{\alpha}Q$ when two terms $P$ and
$Q$ are $\alpha$-convertible.

\begin{figure*}[t] \small
\begin{center}

{\renewcommand\baselinestretch{1.2}\selectfont
\begin{tabular}{@{\sf}rrll@{}} 

$P, Q, R$ & $\Coloneqq$ & & processes\\
& & 0 & null process\\
& $\big|$\ \ & $X$ & process variable\\
& $\big|$\ \ & $(\nu n)P$ & name restriction\\
& $\big|$\ \ & $P\ |\ Q$ & parallel composition\\
& $\big|$\ \ & $P\ ; Q$ & process sequence\\
& $\big|$\ \ & $l[P]$ & located process\\
& $\big|$\ \ & $\bar{a}\langle n\rangle.Q$ & name output\\
& $\big|$\ \ & $\bar{a}\langle P\rangle.Q$ & message output\\
& $\big|$\ \ & $\xi*P$ & input process\\
& $\big|$\ \ & $\overline{\tt{up}}\langle l, P\rangle$ & update provision\\
& $\big|$\ \ & $\tt{up}$$(l, X)\# Q\diamond P$ & update reception\\
& $\big|$\ \ & $\llbracket P\rrbracket$ & blocked process\\

$\xi$ & $\Coloneqq$ & $a(\tilde{x})\ |\ a(\tilde{X})\ |\ l[\tilde{X}]$ & input pattern\\
$*$ & $\Coloneqq$ & $\triangleright\ |\ \diamond$ & \\

$\delta$ & $\Coloneqq$ & $\emptyset$ & null state\\
& $\big|$\ \ & $\{a\}$ & output action, if $\bar{a}\langle n\rangle$
or $\bar{a}\langle P\rangle$ is a valid process\\
& $\big|$\ \ & $\{l\}$ & output source, if $l[P]$ is a valid process\\
& $\big|$\ \ & $\delta \uplus \delta$ & multiset of states\\
\\

$\mathbb{C}$ & $\Coloneqq$ & $\cdot$~~$\big|$~~$(\nu n)\mathbb{C}$~~
$\big|$~~$\xi*\mathbb{C}$ & process context\\
& $\big|$\ \ & $\mathbb{C}~|~P$~~$\big|$~~$\mathbb{C}\ ; P$\\
& $\big|$\ \ & $l[\mathbb{C}]$~~$\big|$~~$\bar{a}\langle
\mathbb{C}\rangle$
~~$\big|$~~$\llbracket \mathbb{C}\rrbracket$\\
& $\big|$\ \ & $\overline{\tt{up}}\langle l, \mathbb{C}\rangle$
~~$\big|$~~$\tt{up}$$(l, X)\# Q\diamond \mathbb{C}$\\
\\
$\mathbb{E}$ & $\Coloneqq$ & $\cdot$~~$\big|$~~$(\nu
n)\mathbb{E}$~~$\big|$~~P~$|~\mathbb{E}$~~$\big|$~~$l[\mathbb{C}]$ &
execution context

\end{tabular}
\par}
\end{center}

\caption{Processes and states of $update\pi$ calculus.}
\label{syntax}
\end{figure*}

The channel of transmitting the update packages is specified by the
name $\tt{up}$, thus its output process is expressed through the
term $\overline{\tt{up}}\langle l, P\rangle$ in our calculus. For
the receiver of update messages, in which the updates are executed
concretely, we define the pattern parameterized process term
$\tt{up}$$(l, X)\# R\diamond P$ to receive the matched update
package on channel $\tt{up}$, and then to activate and execute the
actual updates. Noteworthily, it is possible that an update falls
failure ascribe to some inadequate checks of compatibility before
update and some accidents during update. Thus we introduce a concept
of update log, which allows to incrementally build the log of
updates when each update execution. In the term $\tt{up}$$(l, X)\#
R\diamond P$, the sub-expression $\# R$, in which $R$ is an
accompanied process, is used to associate to each update a process
that records the trace to be stored upon update message reception.
When the update of a process $P$ incurs a recoverable failure, we
block the execution of $P$. This is modeled through the blocked
process $\llbracket P\rrbracket$ that behaves as $P$ but cannot be
activated until it is restored when the execution of a well-defined
update failure recovery (or rollbacking).

A state $\delta$ is a multiset of names that represents the output
actions or resources visible to the update when it was initiated. In
the following, we write $\delta \uplus \{a\}$ for the multiset
$\delta$ enriched with the name $a$ and $\delta\backslash\delta'$
for the multiset obtained from $\delta$ by removing elements found
in $\delta'$, that is the smallest multiset $\delta''$ such that
$\delta\subseteq\delta'\uplus\delta''$. The symbol $\emptyset$
stands for the empty multiset while $\{a^{n}\}$ is the multiset
composed of exactly $n$ copies of $a$, where $\{a^{0}\}=\emptyset$.
The evaluation of state is $\delta=\{a\}$, which $a$ is an output
action, if $\bar{a}\langle n\rangle$ or $\bar{a}\langle P\rangle$ is
a valid process of the current updated components, and
$\delta=\{l\}$, which $l$ is an output resource, if $l[P]$ is a
valid process in  the current updated components.

\section{Dynamic Component Update in $update\pi$}

\begin{figure*}[t] \small
\centering

\begin{center}

{\renewcommand\baselinestretch{1.0}\selectfont
\[
\frac{l=k \quad \textsf{match}(\delta, \delta') \quad l[P]\ |\ (k[X]
\triangleright k[Q]) : \{l\}\uplus\delta \rightarrow k[Q\{P/X\}] :
\delta'} {\begin{array}{l} \overline{\tt{up}}\langle l, P\rangle\ |\
$\tt{up}$(k, X)\#R\ \diamond k[Q]
: \{l\}\uplus\delta \rightarrow\\
\hspace{2.0em} $\tt{up}$(k, X)\#R \diamond k[Q]\ |\ \llbracket
R\{P/X\}\rrbracket\ |\ k[Q(P/X)] :
\{l\}\uplus\delta'\end{array}}\quad (\mathsf{\textsc{R.Update.Ok}})
\]
\[
\frac{l\neq k \quad k[Q] : \delta \rightarrow k[Q'] :
\delta'}{\begin{array}{l} \overline{\tt{up}}\langle l, P\rangle\
|\ $\tt{up}$(k, X)\#R\diamond \ k[Q] : \{l\}\uplus\delta \rightarrow\\
\hspace{2.0em} \overline{\tt{up}}\langle l, P\rangle\ |\
$\tt{up}$(k, X)\#R\diamond \ k[Q'] : \{l\}\uplus\delta'\end{array}}
\quad (\mathsf{\textsc{R.Update.UnMat}})
\]
\[
\frac{\displaystyle l=k \quad !\textsf{comp}(P, Q)}{\begin{array}{l}
\overline{\tt{up}}\langle l, P\rangle\ |\ $\tt{up}$(k, X)\#R
\diamond \ k[Q]
: \{l\}\uplus\delta \rightarrow \\
\hspace{2.0em} $\tt{up}$(k, X)\#R \diamond \ k[Q] : \delta
\end{array}} \quad (\mathsf{\textsc{R.Update.Rest}})
\]
\[
\frac{\displaystyle l=k \quad !\textsf{match}(\delta,
\delta')}{\begin{array}{l} \overline{\tt{up}}\langle l, P\rangle\ |\
$\tt{up}$(k, X)\#R \diamond \ k[Q]
: \{l\}\uplus\delta \rightarrow \\
\hspace{2.0em} $\tt{up}$(k, X)\#R \diamond \ R\{P/X\}\ |\ \llbracket
k[Q\{P/X\}]\rrbracket : \delta'\end{array}} \quad
(\mathsf{\textsc{R.Update.Fail}})
\]

\par}
\end{center}

\caption{Reduction semantics of safe component updates.}
\label{update}
\end{figure*}

We describe in this section a series of rule on safe component
updates to model general process of dynamic updates and possible
update failures. And the time selection of dynamic update and the
state preservation and transformation between versions also are
enforced in our calculus. We first define formally two affiliated
functions about state matchability and process compatibility as
follows.

\begin{definition}
Two process state $\delta$, $\delta'$ are said to be matchable
(noted by \textsf{match}($\delta$, $\delta'$)), if
$\delta\subseteq\delta'$ which means that $\delta$, $\delta'$ are
same or all output actions visible to the update in set $\delta$ are
also included in the set $\delta'$.
\end{definition}

\begin{definition}
Two processes $P$, $Q$ are said to be compatible (noted by
\textsf{comp}($P$, $Q$)), if all output and input interfaces of the
constituent components in process $P$ are compatible with those in
process $Q$.
\end{definition}

\subsection{Safe Update of Component}

We assumed that the component is abstracted with a located process
which can be a composition of multi-components, thus in the output
term $\overline{\tt{up}}\langle l, P\rangle$ , $l$ is a locality
identifier of the enclosing updatable components, which is defined
by the programmer of update packages. That is, the identifier and
its version information of a to-update component specified through
the name $l$ in the update package. However, $P$ is a process
constant which specifies the concrete contents of the update
package. Accordingly, in the input process term $\tt{up}$$(l, X)\#
R\diamond P$, $l$ is location constant which can includes some
version information, etc., used to specify a safe location of
update. And the operator $\diamond$ means that the safe location is
replicative and can be preserved during a update.

During a component update, the two locality identifiers of output
and input interface for updates will be compared to ascertain
whether there is a pending update for the current component, through
the check of name matchability and version correctness. Specially,
if the version number of provided update package is larger than the
one of current component, which founded on the assumptions of update
in last section (in which we don't consider the backwards updates
but only forwards updates), a safe update is triggered as
illustrated in the rule ($\mathsf{\textsc{R.Update.Ok}}$). Then the
process $P$ transmits the actual update operations to the target
components $k[Q]$, and this is expressed by a capture avoiding
substitution $\{P/X\}$ in rule ($\mathsf{\textsc{R.Update.Ok}}$).
And the states of new and old processes keep consistent during the
update, which are expressed with state matchability
$\textsf{match}(\delta, \delta')$. The sub-expression $\# R$ is used
to associate to each update a process $R\{P/X\}$ that stores the
associated runtime states upon update message reception. If the
update is successful, the blocked process $R\{P/X\}$ will not be
activated to promote the normal execution of program.

On the contrary, in rule ($\mathsf{\textsc{R.Update.UnMat}}$),
because the provided update package is incompatible with the current
component, by which either smaller version number or unmatched
component identifiers, the process $P$ is not activated and the
target components will perform its intrinsical functions as normal.
However, even if the version numbers and component identifiers are
validated, the update will be restrained when the injected process
$P$ and the to-update process $Q$ are incompatible as illustrated in
rule ($\mathsf{\textsc{R.Update.Rest}}$).

We assume that each update is well-defined, can be recovered from
failures. The process $R$ is an accompanied log process to store the
executed actions of an update, which to be backtracked in case the
update fails. To account for possible failures of the update
process, the accompanied process $R$, which records the trace
information of executed update and becomes critical part of the
recovery process of the update., will be activated for backtracking
and recovery to abort invalid updates. The process $R$ has no
activity until a failure occurs, becoming then the recovery process
$R$. When an update action occurs. the associated log process is
stored and becomes part of the recovery process of the update. In
rule ($\mathsf{\textsc{R.Update.Fail}}$), the version numbers and
component identifiers are validated and the injected process $P$ and
to-update process $Q$ are compatible, but the two states of new and
old processes are inconsistent during the update. It means that the
update of process $Q$ incurs a recoverable failure, we block the
execution of $Q\{P/X\}$. This is modeled through the blocked process
$\llbracket Q\{P/X\}\rrbracket$ that behaves as $Q\{P/X\}$ but
cannot be activated until it is restored from a well-defined update
failure recovery (triggered by the recovery process $R\{P/X\}$).

\subsection{Timing of Update}

In our calculus, the process sequence $P\ ;Q$ forces a temporal
order between the two operands: process $Q$ will be activated only
after successful completion of $P$. Through the process sequence,
safe update points can be set in some feasible locations of whole
programs, thus the timing of update is enforced by this temporal
operands of the sequential composition operator. For example, if a
update can occur after the execution of process $P$ and before the
execution of process $Q$, then as process sequence $P\ ; \tt{up}(k,
X)\#R\ \diamond k[Q]$, a safe update point will be assigned to the
process $Q$. During the execution of process $P$, a update
corresponding to this update point will be pended until the process
$Q$ is about to be activated. On the contrary, if the process $Q$ is
not the process which is about to be activated immediately in next
time, then this update will keep pending. By dictating when an
update can occur, makes it easier to understand the states of the
program which an update is applied to. And through this mechanism,
we allows the lazy updates\cite{Boyapati03} which the component is
not updated until the to-update process is about to execute.

\subsection{State Transformation}

State transformation is meaningful to map a state of the old
application to a state of the new one. In our calculus, a state
$\delta$ is a multiset of names that represents the output actions
visible to the update when it was initiated. At the time of an
update, a well-defined state transformation function is executed
from component entry points. Then the state $\delta$ recorded in the
updated process (the state at the initiation of the update) will be
compared with the current state $\delta'$ (the mapped state when the
update ends) to check if the update have concurrently made changes
to the updated components for consistency. If the new state
$\delta'$ is consistent with the old one $\delta$ (that is, they
satisfy the Definition 1, denoted as $\textsf{match}(\delta,
\delta')$), the updated component will be able to preserves the old
execution. In other words, the component update is successful as
illustrated the rule ($\mathsf{\textsc{R.Update.Ok}}$). Otherwise,
if the two states $\delta$ and $\delta'$ are verified to be
inconsistent (denoted as $!\textsf{match}(\delta, \delta')$), as in
the rule ($\mathsf{\textsc{R.Update.Fail}}$), the update will fall
into be fail to activate the update recovery process $R\{P/X\}$).

\section{Operational Semantics}

The operational semantics of a process algebra is traditionally
given in terms of a labelled transition system describing the
possible evolution of a process. In general, it is not easy to
define directly a labelled transition system. The manipulation of
names and the side conditions in the rules are non-trivial. On the
other side, if the reduction system is available, the corresponding
labelled transition system can be found. Furthermore, by showing the
correspondence of both the reduction system and the labelled
transition system it is possible to prove the correctness of the
latter.

\subsection{Structure Congruence}

The structural congruence relation, written by $\equiv$, equates all
processes we will never want to distinguish for any semantic reason.

\begin{definition}
Structure congruence $\equiv$ is the smallest equivalence relation
on execution process that satisfies the $\alpha$-conversion law, and
the axioms given in Figure \ref{congrunce}.
\end{definition}

\begin{figure*}[t] \small
\centering

\[
P\ |\ Q \equiv Q\ |\ P ~~~ (\textsc{S.Par.C}) \hspace{3.0em} P\ |\
(Q\ |\ R) \equiv (P\ |\ Q)\ |\ R ~~~ (\textsc{S.Par.A})
\]
\[
P\ |\ 0 \equiv P ~~~ (\textsc{S.Par.N}) \hspace{3.0em} P\ ;(Q\ ;R)
\equiv (P\ ;Q)\ ;R ~~~ (\textsc{S.Seq.A})
\]
\[
0\ ;P \equiv P ~~~ (\textsc{S.Seq.N})\hspace{3.0em} (\nu x)0 \equiv
0 ~~~ (\textsc{S.Nu.Nil}) \hspace{3.0em} \llbracket 0\rrbracket
\equiv 0 ~~~ (\textsc{S.Blk.Nil})
\]
\[
(\nu x)(\nu y)P \equiv (\nu y)(\nu x)P ~~~ (\textsc{S.Nu.C})
\hspace{3.0em} [[(\nu x)P]]\equiv (\nu x)\llbracket P\rrbracket ~~~
(\textsc{S.Nu.Blk})
\]
\[
\frac{\displaystyle x\notin \mathsf{fn}(P)}{\displaystyle P\ |\ (\nu
x)Q \equiv (\nu x)(P\ |\ Q)} ~~~ (\textsc{S.Nu.ParR}) \hspace{3.0em}
\frac{\displaystyle x\notin \mathsf{fn}(Q)}{\displaystyle (\nu x)P\
|\ Q \equiv (\nu x)(P\ |\ Q)} ~~~ (\textsc{S.Nu.ParL})
\]
\[
\frac{\displaystyle x\notin \mathsf{fn}(P)}{\displaystyle P\ ;(\nu
x)Q \equiv (\nu x)(P\ ;Q)} ~~~ (\textsc{S.Nu.SeqS}) \hspace{3.0em}
\frac{\displaystyle x\notin \mathsf{fn}(Q)}{\displaystyle (\nu x)P\
;Q \equiv (\nu x)(P\ ;Q)} ~~~ (\textsc{S.Nu.SeqP})
\]
\[
\frac{\displaystyle a\notin \mathsf{bn}(Q)}{\displaystyle
(\bar{a}\langle P\rangle ~|~R)~;Q \equiv \bar{a}\langle P\rangle
~|~R~;Q} ~~~ (\textsc{S.Seq.Out}) \hspace{3.0em} \frac{\displaystyle
a\notin \mathsf{bn}(Q)}{\displaystyle (\bar{a}\langle P\rangle
~|~R)~|~Q \equiv \bar{a}\langle P\rangle ~|~R~|~Q} ~~~
(\textsc{S.Par.Out})
\]
\[
\frac{\xi \equiv \zeta}{\xi * P \equiv \zeta * P} ~~~
(\textsc{S.Ptn.In}) \hspace{3.0em} \frac{l \equiv l' \quad Q \equiv
Q' \quad P \equiv P' \quad X\in \mathsf{fv}(P)\cap
\mathsf{fv}(P')}{\tt{up}(l, X)\# Q\diamond P \equiv \tt{up}(l', X)\#
Q'\diamond P'} ~~~ (\textsc{S.Update.Rev})
\]
\[
\frac{P =_{\alpha} Q}{P \equiv Q} ~~~ (\textsc{S.Alpha})
\hspace{3.0em} \frac{\displaystyle P \equiv Q}{\displaystyle
\mathbb{C}\{P\} \equiv \mathbb{C}\{Q\}} ~~~ (\textsc{S.Context})
\hspace{3.0em} \mathbb{E}\{\llbracket P\rrbracket\} \equiv
\mathbb{E}\{P\} ~~~ (\textsc{S.Exec.Blk})
\]

\caption{structure congruence relations of processes.}
\label{congrunce}
\end{figure*}

Noted that we do not allow $0$ as a successive neutral element, as
in rule $P\ ;0 \equiv P$, because after the execution of process
$P$, its successor split with a sequence operator $;$ is not always
executed. Because the blocked process $\llbracket P\rrbracket$
behaves as $P$ but cannot be activated until a well-defined update
failure recovery is executed, the $\nu$ restriction operation, as in
rule $(\textsc{S.Nu.Blk})$, can extrude to the top level of a block.
And it behaves as $P$ when a blocked process $\llbracket
P\rrbracket$ is activated, so we use the rule
$(\textsc{S.Exec.Blk})$ to indicate that the execution of process
$\llbracket P\rrbracket$ and $P$ are not distinguishable. And it
should be clarified that besides the process output itself, the
output $\bar{a}\langle P\rangle$ also delegate name output
$\bar{a}\langle n\rangle$, located process output $l[P]$ and update
provision $\overline{\tt{up}}~\langle l, P\rangle$ in this rule.

Notice also that the Ambient-like rule $l[(\nu x)P]*Q \equiv (\nu
x)l[P]*Q$ and $l[(\nu x)P] \equiv (\nu x)l[P]$ are not allowed when
$x \notin \{l\} \cup \mathsf{fn}(Q)$ to enable the safety of
programmable locality where the resources are not extruded outside
their owner. Furthermore, in rule $(\textsc{S.Ptn.In})$,
$(\textsc{S.Update.Rev})$, $(\textsc{S.Alpha})$, and
$(\textsc{S.Context})$, we rely respectively on the structural
congruence relation on input patterns, update reception,
$\alpha$-conversion, and process context.

\subsection{Reduction Semantics}

Figure \ref{reduction} gives the reduction semantics of processes in
$update\pi$ calculus. A reduction is of the form
$P:\delta\rightarrow P':\delta'$ where $\delta$ is the state of
process $P$ and $\delta'$ is an evolution of $\delta$. For
simplicity, the state $\delta$ only records the names of all output
actions visible to $P$ when reduction happens. It grows when an
output representing an internal processing result is produced along
with the execution of processes, and shrinks when an output or
resource is consumed, as illustrated in rule
$(\textsc{R.Out.Name})$, $(\textsc{R.Out.Proc})$,
$(\textsc{R.Out.Pass})$, $(\textsc{R.Update.Prv})$,
$(\textsc{R.In.Name})$, $(\textsc{R.In.Proc})$ and
$(\textsc{R.In.Pass})$ where the process constant $P$ has state
$\delta$.

\begin{figure*}[t]\small
\centering

\[
\bar{a}\langle n\rangle : \{a\} \rightarrow 0 : \emptyset ~~~
(\textsc{R.Out.Name}) \hspace{3.0em} \bar{a}\langle P\rangle : \{a\}
\uplus \delta \rightarrow 0 : \emptyset ~~~ (\textsc{R.Out.Proc})
\]
\[
l[P]: \{l\}\uplus \delta \rightarrow 0 : \emptyset ~~~
(\textsc{R.Out.Pass}) \hspace{3.0em} \overline{\tt{up}}~\langle l,
P\rangle : \{l\}\uplus \delta \rightarrow 0 : \emptyset ~~~
(\textsc{R.Update.Prv})
\]
\[
(\bar{a}\langle n\rangle\ |\ a(x) \triangleright P) :
\delta\uplus\{a\} \rightarrow P\{n/x\} : \delta ~~~
(\textsc{R.In.Name})
\]
\[
\frac{\mathrm{bn}(X)\cap \mathrm{fn}(P)=\phi}{(\bar{a}\langle
P\rangle\ |\ a(X) \triangleright Q) : \delta\uplus\{a\} \rightarrow
Q\{P/X\} : \delta} ~~~ (\textsc{R.In.Proc})
\]
\[
\frac{\mathrm{bn}(X)\cap \mathrm{fn}(P)=\phi}{(l[P]\ |\ l[X]
\triangleright Q)) : \delta\uplus\{l\} \rightarrow Q\{P/X\} :
\delta} ~~~ (\textsc{R.In.Pass})
\]
\[
\frac{P : \delta_{1}\uplus\delta \rightarrow P' : \delta_{1} \quad Q
: \delta_{2} \rightarrow Q' : \delta_{2}}{P\ |\ Q :
\delta_{1}\uplus\delta_{2}\uplus\delta \rightarrow P'\ |\ Q' :
\delta_{1}\uplus\delta_{2}} ~~~(\mathsf{\textsc{R.Comm}})
\]
\[
\frac{P : \delta_{1} \rightarrow P' : \delta_{1}'}{P\ |\ Q :
\delta_{1}\uplus\delta_{2} \rightarrow P'\ |\ Q :
\delta_{1}'\uplus\delta_{2}}
~~~(\mathsf{\textsc{R.Par.L}})\hspace{1.5em} \frac{Q : \delta_{2}
\rightarrow Q' : \delta_{2}'}{P\ |\ Q : \delta_{1}\uplus\delta_{2}
\rightarrow P\ |\ Q' : \delta_{1}\uplus\delta_{2}'} ~~~
(\mathsf{\textsc{R.Par.R}})
\]
\[
\frac{P : \delta_{1} \rightarrow P' : \delta_{1}'}{P\ ;\ Q :
\delta_{1} \rightarrow P'\ ;\ Q : \delta_{1}'}
~~~(\mathsf{\textsc{R.Seq.Fst}})\hspace{2.0em} \frac{P : \delta_{1}
\rightarrow P' : \delta_{1}' \quad Q : \delta_{2} \rightarrow Q' :
\delta_{2}'}{P\ ;\ Q : \delta_{1}\uplus\delta_{2} \rightarrow P'\ ;\
Q' : \delta_{1}'\uplus\delta_{2}'} ~~~
(\mathsf{\textsc{R.Seq.Both}})
\]
\[
\frac{P :\delta\rightarrow P':\delta'}{l[(\nu x)P]:\delta\uplus\{l\}
\rightarrow l[(\nu x) P')]:\delta'\uplus\{l\}}
~~~(\mathsf{\textsc{R.Res}}) \hspace{3.0em}
\frac{P:\delta\rightarrow P':\delta'}{[[P]]:\delta\rightarrow
[[P']]:\delta'} ~~~(\mathsf{\textsc{R.Blk}})
\]
\[
\frac{P \equiv P'\quad \mathrm{match}(\delta_{1}, \delta_{1}') \quad
P' : \delta_{1}' \rightarrow Q' : \delta_{2}' \quad Q' \equiv Q
\quad \mathrm{match}(\delta_{2}', \delta_{2})}{P : \delta_{1}
\rightarrow Q : \delta_{2}} ~~~ (\mathsf{\textsc{R.Eqv}})
\]
\[
\frac{P =_{\alpha} P'\quad \mathrm{match}(\delta_{1}, \delta_{1}')
\quad P' : \delta_{1}' \rightarrow Q : \delta_{2}}{P : \delta_{1}
\rightarrow Q : \delta_{2}} ~~~ (\mathsf{\textsc{R.Alpha}})
\hspace{2.0em} \frac{P : \delta_{1} \rightarrow Q :
\delta_{2}}{\mathbb{E}\{P\} : \delta_{1} \rightarrow \mathbb{E}\{Q\}
: \delta_{2}} ~~~ (\mathsf{\textsc{R.Exec}})
\]

\caption{Reduction semantics for processes and states.}
\label{reduction}
\end{figure*}

Specially, the shrinkage semantics of state are also reflected with
rules $(\textsc{R.Update.Ok})$, $(\textsc{R.Update.Rest})$,
$(\textsc{R.Update.UnMat})$ and $(\textsc{R.Update.Fail})$ in Figure
\ref{update}. And these reduction rules govern the dynamic updates
of components, which allow to recover a failed update since those
stored state information are not discarded. This design choice was
influenced by an expected feature of update handlers: the
possibility to recover updates locally spoiled or completed
failures, where by completed failure we mean a update giving rise to
the recoverable failures of several constituent components in a
application module. This feature would not have problems of
feasibility in a real system, since it could be associated to our
calculus a distributed garbage collection \cite{Veiga05} to remove
recoveries of update no longer essential.

In rule $(\mathsf{\textsc{R.Comm}})$, it can be concerned that the
process $P$ provides some consumable resources, denoted by state
$\delta$, and then the process $Q$ consumes these resources.
However, in process $P~;Q$, besides the sequence in time, there is
no interaction between them. So the evolution of process $P$ and $Q$
can be respectively fulfilled, as in rule
$(\mathsf{\textsc{R.Seq.Fst}})$ and
$(\mathsf{\textsc{R.Seq.Both}})$, where the latter explicitly
expresses temporal order between them. Rule
$(\mathsf{\textsc{R.Res}})$ and $(\mathsf{\textsc{R.Blk}})$ deal
respectively with the initiation of an located process and an
blocked process: an ongoing located resource or process block is
created which holds the newer evaluation state $\delta'$.
Furthermore, in equivalence rule $(\textsc{R.Eqv})$ and
$\alpha$-conversion rule $(\textsc{R.Alpha})$, the two reduction
relations depend on a matchability relation, $\mathrm{match}$, which
associates pairs consisting of two states. As illustrated in
Definition 1, this matchability relation is assumed that define how
a single state matches the other single state of process. The rule
$(\mathsf{\textsc{R.Exec}})$ allows reduction to happen inside
arbitrary execution contexts.

\subsection{Labelled Transition Semantics}

The reduction relation defines the interactive behavior of processes
relative to a context in which they are contained, however, it
covers only a part the behavior of processes, i.e., their local
evolution. In other words, the reduction semantics describes how a
process may interact with another, but not how this process (or
parts of it) may interact with the environment. A labelled
transition system describes these possible intraactions of processes
with the environment. It is easy derive labels from the reduction
semantics given in Figure \ref{reduction}. We first define the
substitution as a (partial) function $\theta :
(\textsf{N}\rightarrow \textsf{N})\uplus (\textsf{V}\rightarrow P)$
from names to names and process variables to $update\pi$ calculus
process. We write $P\theta$ the image under the substitution
$\theta$ of process $P$.

To make a transition means that a process $P$ can evolve into a
process $Q$, and in doing so perform the action $\alpha$. Actions
are given by the grammar in Figure \ref{actions}, where input and
output describe interactions between an agent and its environment,
while a special $\tau$ denotes interaction or silent action. Roughly
speaking, transitions labelled with $\tau$ correspond to the plain
reduction relation. Action $\epsilon$, which $\bar{a}~|~a=\epsilon$,
is introduced to signal the complete match of messages with an input
pattern in a trigger. By definition, the parallel operator $|$ on
actions is associative and commutative, and has $\epsilon$ as a
neutral element. The multiset of actions is defined with the
$\uplus$ operator, i.e. $\alpha\uplus\alpha$, which enforces the
sequential execution of two actions.

\begin{figure*}[t] \small
\centering

$ \alpha ~\Coloneqq ~\epsilon ~~|~~ \tau ~~|~~ a ~~|~~ \bar{a} ~~|~~
\alpha~|~\alpha ~~|~~ \alpha\uplus\alpha $ \caption{Syntax of
actions.}

\label{actions}
\end{figure*}

\begin{figure*}[t] \small
\centering
\[
\bar{a}\langle n\rangle\xrightarrow{\bar{a}\langle n\rangle} 0 ~~~
(\textsc{T.Out.Name}) \hspace{2.0em} \bar{a}\langle
P\rangle\xrightarrow{\bar{a}\langle P\rangle} 0 ~~~
(\textsc{T.Out.Proc}) \hspace{2.0em} l[P]\xrightarrow{l[P]}0 ~~~
(\textsc{T.Out.Pass})
\]
\[
\overline{\tt{up}}~\langle l, P\rangle
\xrightarrow{\overline{\tt{up}}~\langle l, P\rangle} 0 ~~~
(\textsc{T.Update.Prv}) \hspace{3.0em} a(x) \triangleright P
\xrightarrow{a(x)} P\theta : \delta ~~~ (\textsc{T.In.Name})
\]
\[
a(X) \triangleright Q \xrightarrow{a(X)} Q\theta ~~~
(\textsc{T.In.Proc}) \hspace{3.0em} l[X] \triangleright Q
\xrightarrow{l[X]} Q\theta ~~~ (\textsc{T.In.Pass})
\]
\[
\tt{up}(k, X)\#R\ \diamond k[Q] \xrightarrow{\tt{up}(k, X)}
\tt{up}(k, X)\#R \diamond k[Q]\ |\ \llbracket R\theta\rrbracket\ |\
k[Q\theta)]\quad (\mathsf{\textsc{T.Update.Ok}})
\]
\[
\frac{P\xrightarrow{\alpha} P'\quad a\notin
fn(\alpha)}{(\nu{a})P\xrightarrow{\alpha}(\nu{a})P'} ~~~
(\textsc{T.Nu}) \hspace{3.0em}
\frac{P\xrightarrow{\alpha}P'}{[[P]]\xrightarrow{\alpha}[[P']]}
~~~(\textsc{T.Blk}) \hspace{3.0em}
\frac{P\xrightarrow{\tau}Q}{l[P]\xrightarrow{\tau}l[Q]} ~~~
(\textsc{T.Pass})
\]
\[
\frac{P\xrightarrow{\epsilon}Q}{P\xrightarrow{\tau}Q}
~~~(\textsc{T.Red}) \hspace{3.0em}
\frac{P\xrightarrow{\alpha}P'\quad
\alpha\neq\epsilon}{P~|~Q\xrightarrow{\alpha}P'~|~Q} ~~~
(\textsc{T.Par.L}) \hspace{3.0em} \frac{Q\xrightarrow{\alpha}Q'\quad
\alpha\neq\epsilon}{P~|~Q\xrightarrow{\alpha}P~|~Q'} ~~~
(\textsc{T.Par.R})
\]
\[
\frac{P \xrightarrow{\alpha} P'}{P\ ;\ Q \xrightarrow{\alpha} P'\ ;\
Q} ~~~(\mathsf{\textsc{T.Seq.Fst}})\hspace{3.0em} \frac{P
\xrightarrow{\alpha} P' \quad Q \xrightarrow{\beta} Q'}{P\ ;\ Q
\xrightarrow{\alpha\uplus\beta} P'\ ;\ Q'} ~~~
(\mathsf{\textsc{T.Seq.Both}})
\]
\[
\frac{P \xrightarrow{\alpha} P' \quad Q \xrightarrow{\beta} Q' \quad
\alpha\neq\epsilon}{P\ |\ Q \xrightarrow{\alpha~|~\beta} P'\ |\ Q'}
~~~(\mathsf{\textsc{T.Comm}}) \hspace{3.0em} \frac{P
\xrightarrow{\alpha} P'\quad x\notin fn(\alpha)}{l[(\nu x)P]
\xrightarrow{\alpha} l[(\nu x) P')]} ~~~(\mathsf{\textsc{T.Res}})
\]
\[
\frac{P \equiv P' \quad P' \xrightarrow{\alpha} Q' \quad Q' \equiv
Q}{P \xrightarrow{\alpha} Q} ~~~ (\mathsf{\textsc{T.Eqv}})
\hspace{3.0em} \frac{P =_{\alpha} P' \quad P' \xrightarrow{\alpha}
Q}{P \xrightarrow{\alpha} Q} ~~~ (\mathsf{\textsc{T.Alpha}})
\]
\caption{Labelled transition system semantics for processes.}
\label{labeledsemantics}
\end{figure*}

In LTS semantics in Figure \ref{labeledsemantics}, one first can
note that in the labelled transition relation (in Figure
\ref{labeledsemantics}), $\bar{a}\langle n\rangle$, $\bar{a}\langle
P\rangle$, $l[P]$ and $\overline{\tt{up}}~\langle l, P\rangle$ all
give rise to a out transition which respectively has $\bar{a}$,
$\bar{a}$, $l$ and $\overline{\tt{up}}$ as channel name. Specially,
no continuation is to afflicted each output action because we
consider the output process is parallel to other processes, so all
of these out transition will evolve out processes to $0$. Second,
all types of input actions (e.g., $a(x)$, $a(X)$ and $\tt{up}(k,
X)$) or located resources (e.g., $l[X]$) will lead the bound names
of an objective process to be substituted by the names in $x$ or
processes in $X$. Specially, in rule
$(\mathsf{\textsc{T.Update.Ok}})$, the process $Q$ in location $k$
is triggered to execute some well-defined update operations, noted
by the substitution process $Q\theta$, and the accompanied process
$R$ is evolved to a blocked process $R\theta$. Third, the rules
$(\textsc{T.Nu})$, $(\textsc{T.Blk})$, $(\textsc{T.Par.L})$,
$(\textsc{T.Par.R})$, $(\mathsf{\textsc{T.Seq.Fst}})$ and
$(\mathsf{\textsc{T.Res}})$signal that the transition of a process
is not affected by $\nu$ restriction, block restriction, parallel or
sequence operation and passivation. But it is also noted that the
evolution of successive process must be triggered after a successful
execution of its antecedent process, as in rule
$(\mathsf{\textsc{T.Seq.Both}})$, where an explicitly temporal order
is expressed between them.

\section{Related Work}

In general, the systems being updated dynamically are typically
safety-critical, so it is important to select suitable timing of
update to enable correct evolution of updated system. To make sure
that the update is not performed while executing a specific code
region, update authors can specify safe update points
\cite{Neamtiu08} or mark blocks of code that must be entirely
executed on a single version of the system \cite{Neamtiu082}, as
described in our update system which provides formal semantics to
reduce safe update points. While these approaches have been
effective in some real-world scenarios, they potentially suffer from
a structural problem. Update safety constraints are hard-coded in
the original version of the system and cannot be modified at update
time. Thus in this paper, we provide formal semantics to reduce safe
update point.

In \cite{Bierman03}, Bierman et al. proposed a typed
$\lambda$-calculus, named by Update, to reduce dynamic software
update, which multiple versions of a software component can co-exist
in system. This method is relatively intuitionistic but difficult to
implement owing to the requirement of multi-version coexist at one
time. Similar to the idea in our method, Stoyle et
al.\cite{Stoyle07} developed a formal Proteus calculus to model
dynamic update C-like programs, which assumed a new version has
special signature different from its old version and all of updates
are provably safe and consistent. Different from our method, these
existing approaches have largely neglected some other important
aspect, such as state preservation and transformation, possible
update failure and recovery, while focusing more on the executing
process of update.

The ground ideas on our calculus inherit from some extended
higher-order process calculi, e.g. M-calculus\cite{Schmitt03} and
Kell-calculus\cite{Schmitt05}, and specially the component-based
software paradigm is represented with passivation which is similar
to Kell-calculus. In complex systems, different updates may require
very different conditions to be applied. When considering several
categories of updates, the notion of transactional version
consistency\cite{Stoyle07} can be generalized throughout the entire
lifetime of a software system. In \cite{Vaz08}, the authors
presented a calculus for modeling long running transactions within
the framework of the $\pi$-calculus, with support for dynamic
compensation as a recovery mechanism. In our calculus, the failure
recovery of an update apply a similar mechanism to this dynamic
compensation.

\section{Conclusion and Future Work}

In this paper, we propose a dynamic update calculus, $update\pi$
calculus, based on extended higher order process calculi to model
dynamic update of component-based software. The $update\pi$ calculus
is an attempt to extend the higher-order $\pi$ calculus with
passivation, and to enhance it, through the introduction of a family
of dynamic component update mechanisms. The calculus focuses on the
following main concepts: the feasible granularity of update, timing
selection of dynamic update, state transformation between versions,
possible update failures and recoveries. We describe a series of
rule on safe component updates to model some general processes of
dynamic update and discuss its reduction semantics coincides with a
labelled transition system semantics that illustrate the expressive
power of these calculi.

In general, state mapping problem is undecidable\cite{Gupta96}, but
this does not mean that there is no any possibility to solve this
problem automatically or semi-automatically\cite{Bazzi09}. In future
works, we will manage to propose approaches to make us closer to
achieve more practical and reductive state preservation and
transformation. Furthermore, in a component-based system, it is
relatively probable to cooperatively update several constituent
components and some relative problems on this have been researched.
So we will attempt to implement the cooperative update mechanism in
future version of $update\pi$ calculus. And also we will study the
notions of bisimilarity and equivalence for the $update\pi$
calculus, and discuss and prove some corresponding conclusions.

\section*{Acknowledgements}

This paper is partially supported by the National Natural Science
Foundation of China (NSFC) under Grant No.60673116, 60970010, the
National Grand Fundamental Research 973 Program of China under Grant
No.2009CB320705, and the Specialized Research Fund for the Doctoral
Program of Higher Education of China under Grant No.20090073110026.

\end{document}